\documentclass[10pt]{iopart}
\usepackage{latexsym,epsfig,graphicx,bbm}

\newcommand{\ket}[1]{\left | \, #1 \right \rangle}

\newcommand{\bracket}[3]{\left\langle #1 \left| #2 \right| #3 \right\rangle}

\newcommand{\av}[1]{\langle #1\rangle}

\newcommand{\NAT}{{\em Nature }}

\newcommand{\eqr}[1]{equation~(\ref{#1})}
\newcommand{\fir}[1]{figure~\ref{#1}}

\begin{document}

\paper[Signatures of the superfluid to Mott-insulator
transition]{Signatures of the superfluid to Mott-insulator
transition in the excitation spectrum of ultracold atoms}

\author{S~R~Clark~and D~Jaksch}
\address{Clarendon Laboratory, University of Oxford, Parks Road, Oxford OX1 3PU, U.K.}
\ead{s.clark@physics.ox.ac.uk}

\begin{abstract}
We present a detailed analysis of the dynamical response of
ultra-cold bosonic atoms in a one-dimensional optical lattice
subjected to a periodic modulation of the lattice depth. Following
the experimental realization by St{\"o}ferle {\em et al}~[Phys.
Rev. Lett. {\bf 92}, 130403 (2004)] we study the excitation
spectrum of the system as revealed by the response of the total
energy as a function of the modulation frequency $\Omega$. By
using the Time Evolving Block Decimation algorithm, we are able to
simulate one-dimensional systems comparable in size to those in
the experiment, with harmonic trapping and across many lattice
depths ranging from the Mott-insulator to the superfluid regime.
Our results produce many of the features seen in the experiment,
namely a broad response in the superfluid regime, and narrow
discrete resonances in the Mott-insulator regime. We identify
several signatures of the superfluid-Mott insulator transition
that are manifested in the spectrum as it evolves from one limit
to the other.
\end{abstract}


%\tableofcontents

\maketitle

\section{Introduction}
The realization of ultracold atoms confined in optical lattices
has made a large range of fundamental equilibrium and dynamical
phenomena of degenerate quantum gases experimentally accessible.
The success of this approach stems from the fact that, in contrast
to analogous condensed matter systems, optical lattices form a
defect free lattice potential which can trap a dense cloud of
atoms with long decoherence times and can be controlled rapidly
with a great deal of flexibility. This has already enabled a
seminal demonstration of the superfluid (SF) to Mott insulator
(MI) transition by Greiner {\em et al} \cite{Greiner} that was
predicted to occur for a clean realization of the Bose-Hubbard
model (BHM) \cite{Fisher,Jaksch98}. More recently, other themes of
ultracold-atom research have been explored experimentally such as
the purely one-dimensional (1D) Tonks-Girardeau limit
\cite{Paredes,Kinoshita}, the characterization of the SF-MI
transition via the excitation spectrum through 1D - 3D
dimensionality crossover \cite{Stoferle,Kohl}, and impurity
effects caused by Bose-Fermi mixtures \cite{Gunter,Ospelkaus}.

Here we focus on features of the excitation spectrum from
reference \cite{Stoferle} which were revealed for a 1D BHM by
periodic modulation of the lattice depth. The experimental
accessibility of the excitation spectrum provides a rich source of
additional information that can be compared with well studied
quantities such as the dynamic structure factor
\cite{Roth,Batrouni05}. In addition to this the experiment
demonstrates the transition through the evolution of the spectrum
from discrete sharp resonances in the MI regime to a broad
continuum of excitations in the SF regime. Changes in the
structure of the excitation spectrum provide important evidence
for the transition beyond the loss and revival of phase coherence
when ramping the lattice \cite{Greiner}, and can also be used to
diagnose the temperature of the system \cite{Rey,Reischl}. The 1D
system is of particular interest for several reasons. Firstly,
quantum fluctuations are expected to play a strong role there
\cite{Giamarchi}, and this is indeed found to be the case with the
critical ratio of the on-site interaction to kinetic energy
$(U/J)_c$, identified by the appearance of the discrete structure
in the spectrum, being lower than that predicted by mean field
theory \cite{Stoferle}. Secondly, the behavior found in the 1D
experiment for the SF regime, specifically a large and broad
non-zero response, most strikingly departs from standard
theoretical predictions. Specifically, linear response using
Bogoliubov theory for a Bose-Einstein condensate (BEC) in a
shallow 1D optical lattice predicts that lattice modulation cannot
excite the gas in the SF regime due to the phonon nature of the
excitation spectrum \cite{Menotti}. Since the quantum depletion of
the SF in the experiment was significant ($\approx 50$\%), it has
been suggested that this was responsible for the response
\cite{Buchler}. Indeed, it has been shown that only a small amount
of seed depletion is required for non-linear effects like the
parametric amplification of Bogoliubov modes to reproduce the SF
response \cite{Kramer,Tozzo}. More recently still, the use of the
sine-Gordon model and bosonization method has demonstrated that
linear response is non-zero at low frequencies \cite{Iucci}.
Lastly, the study of the 1D system permits the use of quasi-exact
numerical methods, such as the Time Evolving Block Decimation
(TEBD) algorithm, where the fully time-dependent dynamical
evolution of the system can be computed efficiently for systems of
equivalent size to those in the experiment.

In addition to the many-body physics perspective, understanding
the excitation spectrum revealed in \cite{Stoferle} is important
for potential applications of the MI state. The zero particle
number fluctuations for an ideal commensurate MI state make them
attractive candidates for several applications, most notably as a
quantum memory, a basis for quantum
computing~\cite{Mandel,Jaksch99,Dorner,Pachos,Brennen,Clark05},
and quantum simulations of many-body quantum
systems~\cite{Jane,Molmer}. A well understood excitation spectrum
can give valuable information about the nature and stability of
the experimental approximation of the ideal MI state to external
perturbations.

In this paper we study the dynamics of the 1D BHM under lattice
modulation and generate excitation spectra for box and harmonic
trapping of a large system over numerous lattice depths ranging
from the SF to MI regime. We find that much of the features of the
box system can be understood from a small exact calculation.
However, the large system calculations were crucial for the
investigation of signatures in the spectrum which indicate the
transition from MI to SF regime in both trappings. We find that
for a harmonically trapped system with less than unit filling the
spectrum is similar to the commensurately filled box, but with the
transition producing less pronounced signatures. Additional
calculations progressing from the MI regime for a harmonically
trapped system with a central filling greater than unity produces
a spectra that has very good qualitatively agreement with the
experiment. In this way our results are different but
complementary to a very recent study of the same experiment by
Kollath {\em et al} \cite{Kollath06} where they discovered that
excitation spectra can reveal information about the
commensurateness of the system.

The structure of the paper is as follows: we give an overview of
the physical setup for the 1D system in section \ref{model}
followed by a description of the excitation scheme in section
\ref{excite}. We then introduce the linear response formalism for
this scheme in section \ref{lineartheory}, and in section
\ref{numerics} we give an overview of the literature describing
the TEBD simulation method used here for the larger systems. The
results are then presented in section \ref{results}, firstly for a
small box system computed exactly in section \ref{small}, then for
larger systems computed with the TEBD algorithm for box and
harmonic trapping in section \ref{large} and \ref{trapped}
respectively. We then end with the conclusions in section
\ref{conc}.

\section{Probing the system}

\subsection{Optical lattices and the Bose-Hubbard model}
\label{model} In the experiment \cite{Stoferle} effective 1D
systems were formed from an anisotropic 3D optical lattice loaded
with ultra-cold bosonic atoms \cite{Jaksch98}. This is done by
adiabatically exposing a BEC to far-off resonance standing wave
laser fields in three orthogonal directions forming a 3D optical
lattice potential $V_{OL}({\bf r})=\sum_{d=1}^{3}V_{d0}\sin
^{2}(q_Br_d)$ where $q_B=2\pi /\lambda $ and $\lambda $ is the
wavelength of the laser light yielding a lattice period $a=\lambda
/2$ \cite{Jaksch04}.  The height of the potential $V_{d0}$ is
proportional to the intensity of the $d$-th pair of laser beams,
and is conveniently expressed in terms of the recoil energy $E_r =
q_B^2/2m$ for atoms of mass $m$ (taking $\hbar=1$ throughout).

Effective 1D systems are then formed by making laser intensities
in two of the directions $r_2\equiv y$ and $r_3\equiv z$ very
large ($V_{\perp} \approx 30 E_r$). The confinement is then
sufficiently strong to inhibit any tunnelling or excitations in
those directions on the energy scales we are concerned with. The
result is an array of many isolated effective 1D systems in the
$r_1 \equiv x$ direction
~\cite{Stoferle,Moritz,Paredes,Kinoshita}. For the remaining
lattice intensity $V_{10}\equiv V_0$ we consider much shallower
depths, but always remain deep enough ($V_0 > 4 E_r$) to ensure
that there is an appreciable band-gap between the lowest and first
excited Bloch band, given in the harmonic approximation by
$\omega_{ho}=2 E_r\sqrt{V_0/E_r}$. Combined with the ultra-low
temperatures of the atoms this is sufficient to ensure that the
dynamics can be described by the lowest Bloch band of the lattice,
and that the tight-binding approximation is applicable
\cite{Jaksch98}.

With the centre of lattice site $j$ in one such 1D system given by
$x_j = j a$ we can construct a complete and orthonormal set of
localized mode functions $\phi_j({\bf r})=w(x-x_j)W(y)W(z)$
factorized as the product of Wannier functions $w$ and $W$ of the
lowest Bloch band for the shallow and deeply confined directions
respectively. After expanding the bosonic field operator
$\hat{\psi}\left({\bf r}\right)$ into these modes and restricting
our consideration to one 1D system, the resulting Hamiltonian $H$
reduces to the 1D BHM \cite{Jaksch98} composed of $M$ sites
\begin{equation}
H=-J\sum_{j=1}^{M-1}(b_j^{\dagger }b_{j+1}+{\rm
h.c.})+\sum_{j=1}^M v_j b_j^\dagger b_j + \frac{U}{2}\sum_{j=1}^M
b_j^\dagger b_j^\dagger b_j b_j, \label{HBHM}
\end{equation}
where the operators $b_j$ ($b_j^\dagger$) are bosonic destruction
(creation) operators for an atom in site $j$. The parameters of
the BHM are functions of the lattice depth $V_0$ with the matrix
elements for hopping between adjacent sites $j$ and  $j+1$ and
on-site interaction strength given by \cite{Zwerger}
\begin{eqnarray}
J &=& -\int dx~w(x-x_j)\bigg(-\frac{\hbar^2}{2m} \frac{d^2}{d
x^2}+ V_0 \sin^{2}(q_Bx) \bigg)
w(x-x_{j+1}),\nonumber \\
U &=& 2a_{s}\omega_{ho} \int dx ~|w(x-x_j)|^{4}, \label{UJparams}
\end{eqnarray}
where $a_{s}$ is the $s$-wave scattering length, and a Gaussian
ansatz has been used for the tightly confined Wannier states $W$.
The trapping offset is well approximated as $v_j \approx
V_T(x_j,y,z)$, where $V_{T}({\bf r})$ describes an additional
slowly varying trapping potential which could be due to magnetic
trapping. In the case of \cite{Stoferle} the axial potential of
the 1D system was dominated by the Gaussian beam envelopes (with
$1/e^2$ waists $l$) of the lasers for the strongly confined
directions characterized by the trapping frequency $\omega_T =
2E_r\frac{\lambda}{\pi l}\sqrt{V_{\perp}/E_r}$ \cite{Moritz}.

The physics of the BHM is governed by the ratio $U/J$. Competition
between these two terms results in a transition at temperature
$T=0$ for a critical ratio $(U/J)_c$ from the SF to the MI regime
\cite{Fisher,Sachdev}. Mean-field theory for an infinite unit
commensurately filled 1D system predicts that $(U/J)_c \approx 2
\times 5.8$. However, if the strong quantum fluctuations present
in 1D are taken into account, the appearance of the SF regime is
not predicted to occur until the critical ratio drops to $(U/J)_c
\approx 3.85$ \cite{Kuhner}. The presence of trapping and the
finite size of a system modifies the nature of the transition,
prohibiting it from being sharp and so in line with the experiment
\cite{Stoferle} we expect the transition to occur gradually
somewhere in between these limits \cite{Batrouni02}.

\subsection{Lattice modulation excitation scheme}
\label{excite} In the experiment \cite{Stoferle} the 1D system was
initially prepared in the groundstate for some depth $V_0$,
ranging from the SF to MI regime. The axial lattice depth was then
subjected to a modulation of the form
\begin{equation}
V_{OL}(x,t)=V_0[1 + A\sin(\Omega t)]\sin^2(q_Bx), \label{Modulate}
\end{equation}
where $A$ is the modulation amplitude as a fraction of the initial
lattice depth, and $\Omega$ is the modulation
frequency~\cite{Stoferle}. The modulation was applied for a fixed
time $\tau = 30$~ms after which the energy deposited into the
system was measured by time-of-flight imaging of central momentum
width averaged over the many 1D systems realized. The applied
modulation frequency was taken to a maximum of $\Omega/2\pi =
6$~kHz which defines the relevant energy scale for the system and
was well below the band-gap.

To compare with the experiment we take the wavelength of the light
used to form the optical lattice as $\lambda=826$~nm, and the
atomic species trapped as $^{87}$Rb, where $a_s=5.1$~nm, in all
numerical values quoted. For our calculations we initially
computed the groundstate of the system over depths
$U/J=5,6,\dots,20$ for the large systems and slightly shallower
depths $U/J=2,3,\dots,20$ for the small system \cite{Shallow}, all
with fixed particle number $N$. We study a small and large system
with box boundaries which are commensurately filled as $M=N=7$ and
$M=N=41$ respectively. We also consider a harmonically trapped
system where $v_j=m \omega_T^2 x_j^2/2$ for a slightly smaller
system with $M=25$ using $\omega_T/2\pi = 70$ Hz and $N=15$, as
well as $\omega_T/2\pi = 100$ Hz and $N=30$. These trapping
frequencies are close to that in the experiment
\cite{Stoferle,Moritz} where $\omega_T/2 \pi \approx 85$ Hz, and
where sufficient to eliminate any occupation at the box boundaries
of the system. The modulation given in \eqr{Modulate} was then
applied to the system by computing time-dependent BHM parameters
via \eqr{UJparams}. This includes an implicit assumption that the
Wannier states describing atoms in the lattice adiabatically
follow the variations in the lattice potential induced by the
modulation. Given that the timescale of atomic motion in a lattice
site is $\nu=\omega_{ho}/2\pi$ and that this is typically an order
of magnitude greater than the modulation frequencies applied, the
adiabatic assumption is reasonable. The response of the system was
then measured via the total energy $\av{H_0}$, with respects to
the unperturbed BHM, for different $\Omega$. To demonstrate the
evolution of the spectra at different depths we use the same fixed
range for $\Omega/2\pi$ at all depths. This range is identical to
the experiment and is likewise quoted in kHz while the energy
absorbed is expressed in units of $E_r$.

\section{Analysis}
\subsection{Linear response}
\label{lineartheory} We consider a straightforward linear response
treatment of this excitation scheme which follows under the same
assumption used for the numerical calculation that the system is
described by the BHM with time-varying parameters $J[V_{OL}(t)]$
and $U[V_{OL}(t)]$. For perturbative calculations we make an
additional assumption that the modulations are weak and
approximate the variation of these functions about the initial
depth $V_0$ linearly resulting in a harmonic perturbation
\cite{Iucci}
\begin{equation}
H(t) = H_0 + AV_0\sin(\Omega t)\left(\delta U H_0 - J_0\{\delta J
- \delta U\}H_J\right), \label{Perturbation}
\end{equation}
where $H_0=-J_0H_J + U_0H_U$, $\delta U = \frac{d \ln
U}{dV}|_{V_0}$, $\delta J = \frac{d \ln J}{dV}|_{V_0}$, $U_0 =
U[V_0]$ and $J_0 = J[V_0]$. The perturbation is split into a part
that is proportional to the unperturbed BHM Hamiltonian $H_0$ and
a part proportional to the hopping operator $H_J =
\sum_{j=1}^{M-1}(b_j^{\dagger }b_{j+1}+{\rm h.c.})$ under the
proviso that $U_0 > J_0$. The first part cannot induce excitations
and instead gives a small time-dependent shift to the unperturbed
energies which can be ignored. As a result the excitation operator
of this perturbation is just the hopping operator
\cite{Iucci,Reischl} with coupling $\kappa = AV_0J_0(\delta J -
\delta U)$. By acting over the whole system uniformly it creates
excitations with zero quasimomentum as expected.

Let us label the eigenstates of the unperturbed BHM $H_0$ as
$\ket{n}$ with energy $\epsilon_n$. The principle quantities of
interest are the excitation probabilities $P_{0\rightarrow
n}(\tau,\Omega)$ for the transitions to the excited states
$\ket{n}$ from the groundstate $\ket{0}$ due to this perturbation
being applied for a time $\tau$ with a frequency $\Omega$. In
first-order time-dependent perturbation theory these are given by
\begin{equation}
P^{(1)}_{0\rightarrow n}(\tau,\Omega)=\left|t^{(1)}_{0\rightarrow
n}\right|^2 =
\kappa^2\left|\bracket{n}{H_J}{0}I^{(1)}_n(\tau,\Omega)\right|^2
,\label{first}
\end{equation}
where $I^{(1)}_n(\tau,\Omega) = \int_0^\tau
dt~e^{i\omega_{n0}t}\sin(\Omega t)$ and
$\omega_{n0}=(\epsilon_n-\epsilon_0)$. This result reduces under
the rotating wave approximation and the limit $\tau \rightarrow
\infty$ to the familiar Golden rule result. We also make use of
the second-order result
\begin{equation}
P^{(2)}_{0\rightarrow n}(\tau,\Omega)=\left|t^{(1)}_{0\rightarrow
n} - \kappa^2\sum_m
\bracket{n}{H_J}{m}\bracket{m}{H_J}{0}I^{(2)}_{nm}(\tau,\Omega)\right|^2
,\label{second}
\end{equation}
where $I^{(2)}_{nm}(\tau,\Omega)=\int_0^\tau dt~\int_0^t
dt'~e^{i\omega_{nm}t}e^{i\omega_{m0}t'}\sin(\Omega t)\sin(\Omega
t')$. The total energy absorbed by the system, relative to $H_0$,
is then $E(\tau,\Omega) = \sum_n \epsilon_n P_{0\rightarrow
n}(\tau,\Omega) - \epsilon_0$.

\subsection{Numerical method}
\label{numerics} For the large systems investigated later in
sections \ref{large} and \ref{trapped} exact integration of the
many-body Schr{\"o}dinger equation is not feasible. To compute
these results we employed the TEBD algorithm
\cite{Vidal03,Vidal04} which is a quasi-exact numerical method
that allows the dynamical evolution of 1D quantum lattice systems
with nearest-neighbor interactions to be computed efficiently and
accurately. The algorithm has been successfully applied to
numerous physical systems including the BHM
\cite{Clark04,Daley04,Daley05,Gobert,Kollath06,Kollath05}. Not
long after being proposed by Vidal it was recognized
\cite{Daley04,White04} that TEBD shares some conceptual and formal
similarities with the well established density matrix
renormalization group (DMRG) \cite{White92,Schollwock} method
enabling the development of a new adaptive time-dependent-DMRG
algorithm which incorporates optimizations from both. A detailed
analysis of the accuracy and error propagation of this method was
given by Gobert {\em et al} \cite{Gobert} and applies quite
generally to TEBD also. We do not describe the TEBD algorithm
here, except to mention some specific issues, and instead refer
the reader to the relevant articles above for more details.

The objective of the simulations in this paper were to map out the
response of the system with the modulation frequency $\Omega$ and
depths $U/J$ which cross the SF-MI transition. For sufficient
sampling this required in excess of 1000 simulations for both the
box and harmonically trapped systems presented here. Consequently
we were limited for practical reasons to using a truncation
parameter $\chi=30$ (see \cite{Vidal04} for an explanation of
$\chi$) for all simulations which was lower than that strictly
necessary in order to achieve full convergence. Thus we cannot
claim that our calculations are quasi-exact and would not expect
the many-body state given by the simulation at the end to have a
high fidelity with the true state of the system. Instead we treat
our calculation as an approximation in the same spirit as the
Gutzwiller ansatz \cite{Fisher,Krauth} (where $\chi=1$), but with
the important difference that since $\chi>1$ we are permitting a
non-negligible amount of quantum correlations. Given that we are
only interested in the total energy, which is an observable
composed of one- and two-particle correlations, we expect that
this approximation should yield quantitative agreement for the
system sizes and regimes considered here. Indeed, we have carried
out more accurate calculations at specific points which reveal
that the total energy is a robust observable with respects to
truncation \cite{Banuls}. Consequently, the features of the energy
spectrum do emerge, even with this relatively low $\chi$, and the
approximations made do not invalidate the results presented here.
For simulations of the lattice modulation problem at some specific
lattice depths with larger $\chi$ see the recent work by Kollath
{\em et al} \cite{Kollath06}.

\section{Results}
\label{results}

\subsection{Exact calculation for a small system}
\label{small} We begin by studying the exact dynamical evolution
of the BHM under the lattice modulation for a small system
composed of $M=7$ sites and $N=7$ atoms with box boundary
conditions. This system is large enough to produce many of the
essential physical features while still permitting the exact
eigenstates to be computed~\cite{Munzinger}. In fact we find that
much of what is learnt from the small system can be directly
applied to the larger systems. We make use of this by computing
the energy spectrum $\epsilon_n$ and perturbation matrix elements
$\bracket{n}{H_J}{0}$ for two lattice depths with $U/J=4$ and
$U/J=20$ representing the SF and MI regimes respectively. In
\fir{exactspec}(a) the MI spectrum is shown with its
characteristic gapped structure composed of Hubbard bands located
around multiplies of the dominant interaction energy $U$ and
spread by finite hopping $J$. In addition to the spectrum, the
perturbation matrix elements $\bracket{n}{H_J}{0}$ which are of
order $\mathcal{O}(J/U)$ or above are shown as the vertical lines.
For the MI the most numerous (and strongest) contributions are to
the $U$-Hubbard band which is described by 1-particle-hole (1-ph)
excitations like those depicted in \fir{exactspec}(c)(i). The
matrix elements to the $2U$-Hubbard band, which is composed of two
1-ph excitations shown in \fir{exactspec}(c)(ii), cancel to first
order explaining the absence of lines for this manifold. However,
there are a small number of matrix elements to first order
connecting the groundstate to the $3U$-Hubbard band via two 1-ph
excitations with both particles on the same site as in
\fir{exactspec}(c)(iii), but not to three 1-ph excitations as in
\fir{exactspec}(c)(iv). In \fir{exactspec}(b) the `gapless' SF
spectrum is shown. Here in contrast to the MI there are
significant contributions to $\bracket{n}{H_J}{0}$ stretching from
below an energy of $U$ to below $3U$. Separated from this there
are contributions tightly distributed around $4U$. Given the
comparatively equal strengths of the hopping and interaction terms
for the SF regime at $U/J=4$ there is no simple picture of either
the groundstate or the excitations related to these contributions
as there was for the MI regime. Despite this the exact eigenstates
for this small system do reveal two important details: firstly two
1-ph configurations like those in \fir{exactspec}(c)(iii) have an
average energy relative to the $U/J=4$ groundstate which exceeds
$4U$; and secondly these types of configurations are the dominant
contributions in the relevant eigenstates around $4U$. This
overlap in the nature of the excitations points to the
possibility, which we shall shortly confirm, that a resonance to
$3U$ excitations in the MI regime will evolve into a $4U$
resonance as the SF regime is entered.

\begin{figure}[h]
\begin{center}
\includegraphics[width=14cm]{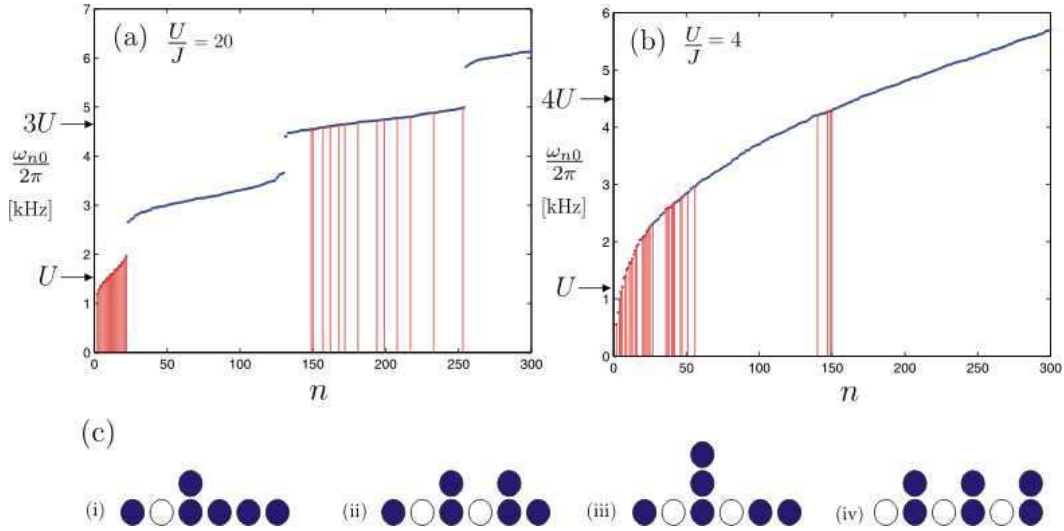}
\caption{The energy spectra for the first 300 eigenstates $n$ for
(a) $U/J=20$ MI regime, and (b) $U/J=4$ SF regime. In both cases
the vertical (red) lines denote the presence of a matrix element
$|\bracket{n}{H_J}{0}|$ connecting an excited state to the
groundstate which is of order $\mathcal{O}(J/U)$ or above. In (c)
a schematic depiction of the excitations which exist in the (i)
$U$-, (ii) $2U$- and (iii-iv) $3U$-Hubbard bands of the MI regime
in (a) is given.}\label{exactspec}
\end{center}
\end{figure}

We first consider the modulation scheme with a weak amplitude
$A=0.01$ where linear-response is applicable. In
\fir{linearresp}~(a) the total energy of the system $\epsilon$ is
shown after the modulation as a function of $\Omega$ for the MI
regime with $U/J=20$. As expected from the matrix elements
$\bracket{n}{H_J}{0}$ in \fir{exactspec}(a) there is strong
response centred around $U$ spread by the width of the $U$-Hubbard
band which is approximately $10J \sim 0.8$~[kHz], and a smaller
(barely visible) response at $3U$. The linear response predictions
based on \eqr{first} are shown also and agree well. For the
largest peak the growth of energy over the modulation time is
shown in \fir{linearresp}(b) and the slight overestimation of the
energy by linear-response is evident at later times. The same
calculation for the SF regime with $U/J=4$ is shown in
\fir{linearresp}(b). Again the results of linear-response agree
well and the structure follows that of perturbation matrix
elements with a broader response between $U$ and $3U$ and an
equivalent strength resonance at $4U$.

\begin{figure}[h]
\begin{center}
\includegraphics[width=14cm]{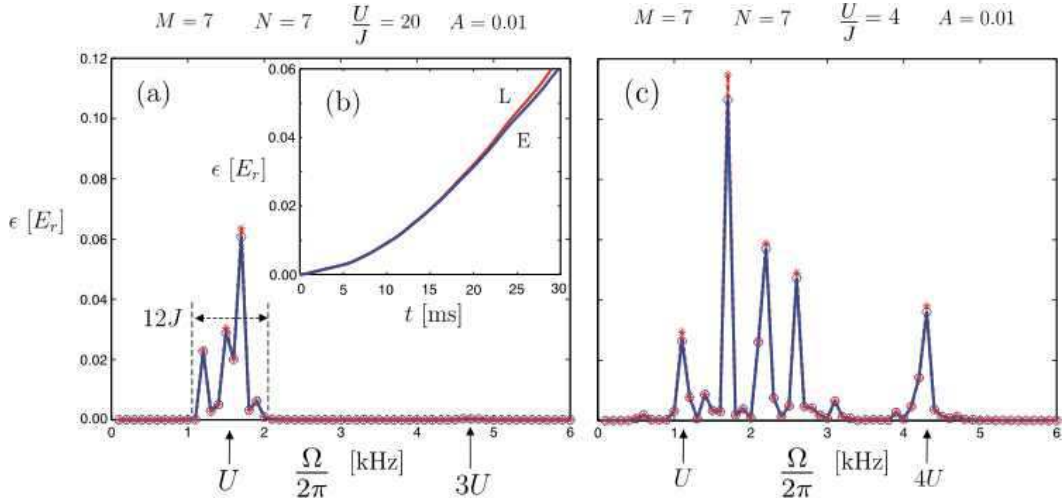}
\caption{For a weak modulation of the small system we have (a) the
total energy $\epsilon$ absorbed in the MI regime with $U/J=20$ as
a function of $\Omega$, (b) $\epsilon$ over time for the largest
MI peak at $\Omega/2\pi \sim 1.7$~[kHz] for the exact (E) and
linear-response (L) calculations, and (c) $\epsilon$ against
$\Omega$ for the SF regime with $U/J=4$. In (a) and (c) the
markers (o) and (*) denote the exact and linear response results
respectively, while the lines are drawn to guide the
eye.}\label{linearresp}
\end{center}
\end{figure}

In line with the experiment~\cite{Stoferle} from this point on we
consider much stronger modulations with $A=0.2$. The result of
these for the MI and SF regimes is given in \fir{smallspec}. For
the MI in \fir{smallspec}(a) we see that the discrete resonances
around $U$ have now filled out into a single peak centred on $U$
that is about 25\% wider than the $U$-Hubbard band at $\sim 1$
[kHz], and that the stronger modulation has now increased height
of the $3U$ peak relative to the $U$ peak. In \fir{smallspec}(a)
we have also included both first- and second-order perturbation
theory results. For such strong modulations the applicability of
perturbation theory is highly questionable, especially for the
long times considered here. This is exemplified by the gross
overestimation of the central peak at $U$ by both linear- and
quadratic-response. In \fir{smallspec}(b) curve (i) shows the
saturation of energy absorption for the central peak at $U$ and
departure from linear response after a short time. In contrast for
the peak at $3U$ \fir{smallspec}(b) curve (ii) shows that
linear-response underestimates the energy absorbed due to its
neglect of the role these eigenstates play in the indirect
processes to higher energies. However, the use of
quadratic-response given by \eqr{second} provides some useful
insight into the additional structure seen for the strong
modulation which do not appear for weak modulations and is not
predicted by linear-response. Specifically, the small satellite
peaks either side of the dominant peak at $U$, which are located
at $U/2$ and $3U/2$ respectively, only appear at second and higher
order leading to their natural identification as the absorption of
two quanta of energy $\Omega$ from the perturbation to reach to
$U$- and $3U$-Hubbard bands. We note also that even under strong
modulations no resonance at $2U$ is seen in agreement with recent
findings~\cite{Kollath06} for commensurate systems. For the SF
regime in \fir{smallspec}(c) a broad response is seen spanning the
region between $U$ and $3U$ consistent with linear response, but
higher-order effects have resulted in saturation and merging of
linear response peaks. As with weaker modulations there is a
separate resonance centred at $4U$ which is now as equivalent in
strength and broader.

\begin{figure}[h]
\begin{center}
\includegraphics[width=14cm]{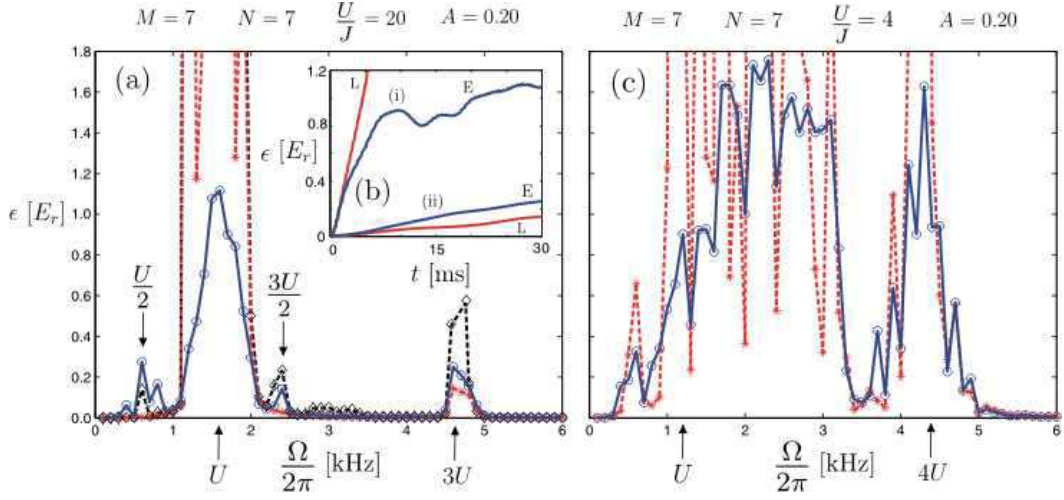}
\caption{For a strong modulation of the small system we have (a)
the total energy $\epsilon$ absorbed in the MI regime with
$U/J=20$ as a function of $\Omega$, (b) $\epsilon$ over the time
at the MI peaks (i) $U$ and (ii) $3U$  for the exact (E) and
linear-response (L) calculations, and (c) $\epsilon$ against
$\Omega$ in the SF regime with $U/J=4$. For (a) and (c) the
markers (o), (*) denote the exact and linear-response results
respectively, while in (a) ($\Diamond$) denotes the
quadratic-response. The lines are drawn to guide the
eye.}\label{smallspec}
\end{center}
\end{figure}

With the full exact calculation we have computed the response of
the system for a sequence of depths $U/J=2,3,\dots,20$ which are
displayed in \fir{smallfull}~(a) demonstrating the evolution of
the MI spectrum in \fir{smallspec}~(a) into SF spectrum in
\fir{smallspec}~(c) with decreasing lattice depth. By examining
the accompanying colour-map in \fir{smallfull}~(b) we can make a
number of initial observations regarding the changing
characteristics of the spectrum over the SF-MI transition.
Firstly, as with the weak modulations, the MI peak at $U$ is seen
to broaden and shift upwards in energy into the SF response, and
this change is most prominent after the depth is lower than $U/J
\sim 12$. Secondly, it is now clear that the resonance at $3U$ for
deep in the MI regime reduces only slightly in energy and ends up
as the $4U$ resonance in the SF regime. Additionally, in line with
weaker modulations in \fir{linearresp}, the $4U$ peak is stronger
than the $3U$ peak it evolves from, and signatures of this change
are already visible in \fir{smallfull}~(a). Finally, by
progressing to slightly shallower depths ($U/J=2,3$) in
\fir{smallfull} we see the broadening of the SF spectrum over the
entire frequency range considered caused by the eventual merging
of the $U$ to $3U$ response with the $4U$ peak.

\begin{figure}[h]
\begin{center}
\includegraphics[width=14cm]{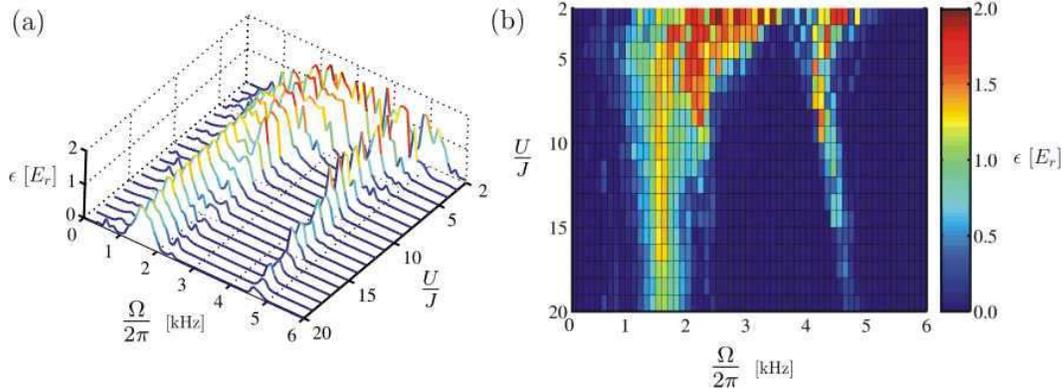}
\caption{(a) The total energy $\epsilon$ absorbed by a small box
system over a sequence of depths, ranging from the MI to SF
regimes, as a function of $\Omega$ for strong modulations. (b) A
colour-map of plot (a).}\label{smallfull}
\end{center}
\end{figure}

\subsection{Large system in a box}
\label{large} To put some of the observations made in the previous
section on a firmer footing we now examine a larger system with
$M=41$ and $N=41$, again with box boundaries, under the same
modulation scheme with $A=0.2$. To compute the dynamical evolution
for this system we resort to the TEBD algorithm using a lattice
site dimension $d_s=5$. In \fir{boxspec} the one-particle density
matrix $\rho_{jk} = \av{b^{\dagger}_jb_k}$ for the groundstate of
(a) the MI with $U/J=20$ and (b) the SF with $U/J=5$ is plotted,
along with the corresponding spectra obtained from these
groundstates in (c) and (d) respectively. As might be expected the
increase in the size of the system reduces finite size effects
resulting in much smoother excitation profiles. However, the
essential features of these plots still follow directly from our
analysis of a small system in terms of both the position and width
of the resonances. Specifically, the width of the $U$ peak in
\fir{boxspec}(c) is again slightly larger than the width of the
$U$-Hubbard band for this system given approximately by first
order perturbation theory as $12J \sim 1$ [kHz]. The effects of
saturation also appear to follow similarly with the maximum energy
absorbed per particle being nearly identical for the two system
sizes.

\begin{figure}[h]
\begin{center}
\includegraphics[width=14cm]{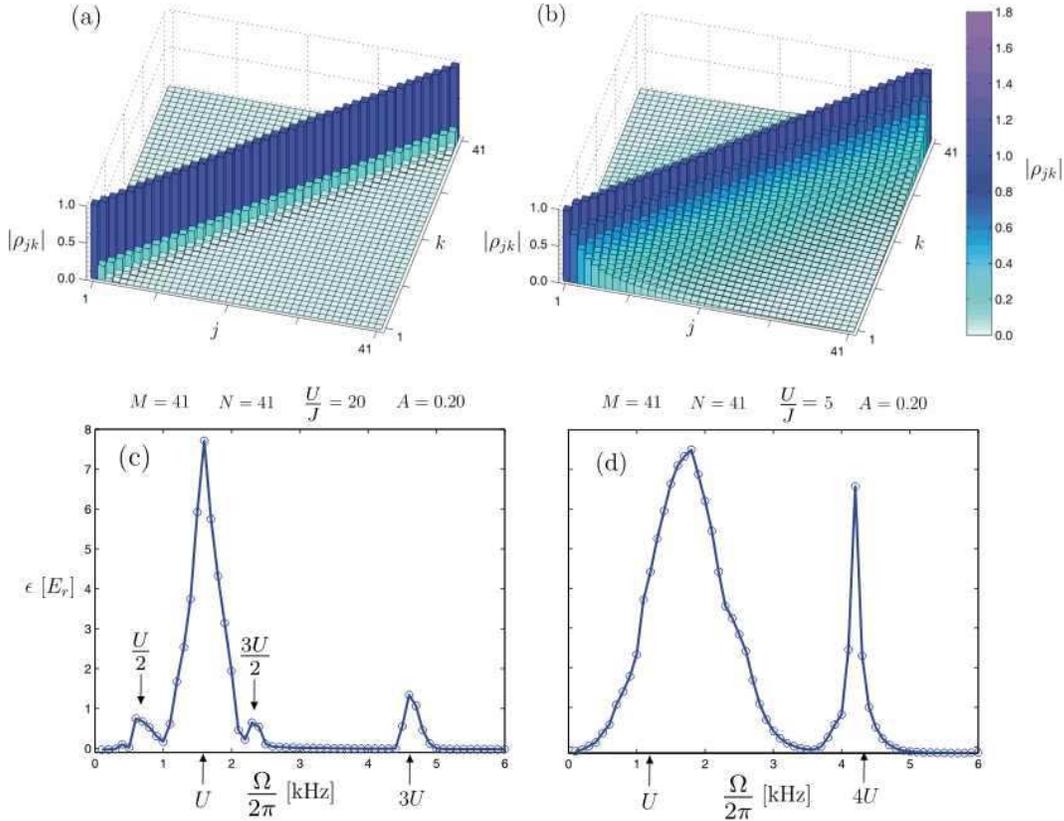}
\caption{The one-particle density matrix $\rho_{jk}$ of the
groundstate for (a) the MI regime with $U/J=20$ and (b) the SF
regime with $U/J=5$. The corresponding excitation spectra of the
total energy $\epsilon$ against $\Omega$ for these groundstate is
shown in (c) and (d) respectively.}\label{boxspec}
\end{center}
\end{figure}

The absorption spectrum over a sequence of depths
$U/J=5,6,\dots,20$ ranging from the SF to MI regime are shown in
\fir{boxfull}. The increased size of the system permits us to
investigate more conclusively some of the observations made
earlier regarding the evolution of the absorption spectra with
decreasing depth. Firstly, we focus on the maximum strength of the
$3U \rightarrow 4U$ peak with $U/J$ which is shown in
\fir{boxanalysis}~(a). The plot shows that the strength of this
peak displays a distinct alteration in its behavior close to $U/J
\sim 10$ where it jumps from an increasing to a decreasing curve.
This behaviour is consistent with what was seen earlier for the
small system and can be attributed to the increasing SF character
with a strong $4U$ peak becoming dominant over the weaker $3U$ MI
peak.

\begin{figure}[h]
\begin{center}
\includegraphics[width=14cm]{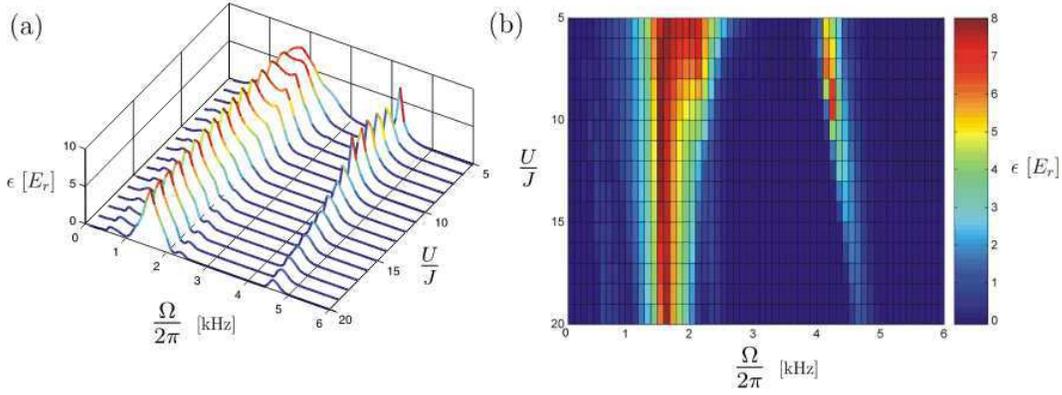}
\caption{(a) The total energy $\epsilon$ absorbed by a large box
system over a sequence of depths, ranging from the MI to SF
regimes, as a function of $\Omega$ for strong modulations. (b) A
colour-map of plot (a).}\label{boxfull}
\end{center}
\end{figure}

\begin{figure}[h]
\begin{center}
\includegraphics[width=14cm]{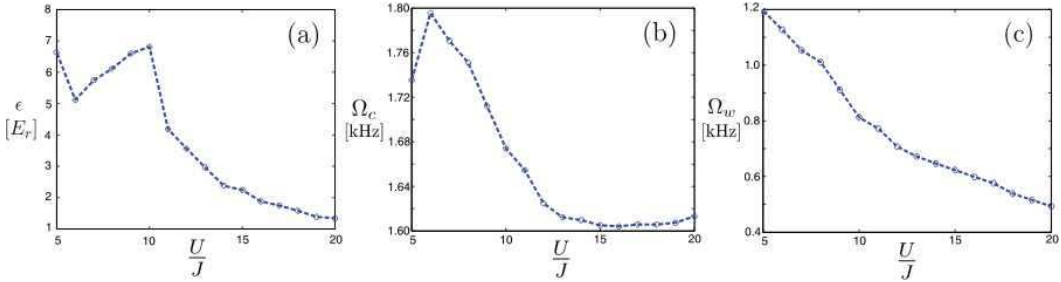}
\caption{(a) The maximum energy $\epsilon$ absorbed for the $3U
\rightarrow 4U$ peak with the lattice depth $U/J$. (b) The centre
position $\Omega_c$, and (c) width $\Omega_w$  of the $U$
resonance with the lattice depth $U/J$. The dotted lines in all
cases is drawn to guide the eye.}\label{boxanalysis}
\end{center}
\end{figure}

To understand the changes exhibited by the resonance around $U$ we
fit this peak with a smoothed-box function of the form
\begin{equation}
\epsilon(\Omega) = \epsilon_{\textrm{min}} +
\mathcal{N}\frac{\epsilon_{\textrm{max}} -
\epsilon_{\textrm{min}}}{1+e^{[(\Omega-\Omega_c)^2-w^2]/s^2}},
\label{box}
\end{equation}
where $\Omega_c$, $w$ and $s$ are parameters specifying the
center, top width and step size of the profile respectively,
whilst $\mathcal{N}=1+e^{-w^2/s^2}$ is the scaling factor. The
advantages of this function (which are even more apparent in the
next section) is that it faithfully describes both the width and
centre location of broad topped resonances. The centre $\Omega_c$
and the full-width-half-maximum $\Omega_w$ extracted from this
fitting then characterizes quantitatively the behaviour which is
evident in \fir{boxfull}~(b).

In \fir{boxanalysis}~(b) we see that the centre $\Omega_c$ of the
$U$ peak remains relatively stationary until $U/J \sim 12$ where
it then shifts more dramatically to higher energies. This is again
a signature of increasing SF character consistent with the
disappearance of the gap between the $U$- and $2U$-Hubbard bands
and the shifting of the strongest contributions to
$\bracket{n}{H_J}{0}$ to higher frequency seen for the small
system. The width $\Omega_w$ in \fir{boxanalysis}~(c) instead
displays a gradual increase without any pronounced changes.

It is clear from the structure of both the MI and SF spectra shown
in \fir{boxspec}(c) and (d) that the box system has some
differences compared to the 1D spectra obtained in the experiment.
These differences are not surprising given that the SF state of
the box system at $U/J=5$ exhibits significant quantum depletion
of $75\%$, which is much larger than that in the experiment. Also,
at all depths considered here the groundstate of the box has a
near homogeneous density of 1 atom per site, as can be seen in
\fir{boxspec}(a) and (b), which is different from the harmonically
trapped system in the experiment.

\subsection{Large harmonically trapped system}
\label{trapped} To approach a setup closer to that of the
experiment we now consider a system with $M=25$ and $N=15$
superimposed with harmonic trapping using $\omega_T/2\pi=70$ Hz. A
significant difference is that at $U/J=5$ the depletion of the SF
state is reduced to $50\%$ in line with the experiment for this
setup. The one-particle density matrices for the MI and SF
groundstates in this case are shown in \fir{harmonicspec}(a) and
(b) respectively. We now see that the MI state is composed of a
central core with unit filling along with small SF lobes at the
edges. This change in structure compared to \fir{boxspec}(a)
introduces additional types of excitations such as those caused by
particles hopping into the vacuum surrounding the MI core. For the
SF state there are now significantly greater off-diagonal
correlations in $\rho_{jk}$ compared to the box system. Despite
these differences, however, the form of the spectrum shown in
\fir{harmonicspec}(c) and (d) remains very close to the box system
displaying the same characteristic peaks as before. The effects of
trapping manifest themselves in this case by flattening of the $U$
peak, along with the broadening and shifting of the $4U$ peak. The
maximum energy per particle again remains approximately equal to
the previous cases considered, showing that the saturation effects
are not significantly altered by the trapping.

\begin{figure}[h]
\begin{center}
\includegraphics[width=14cm]{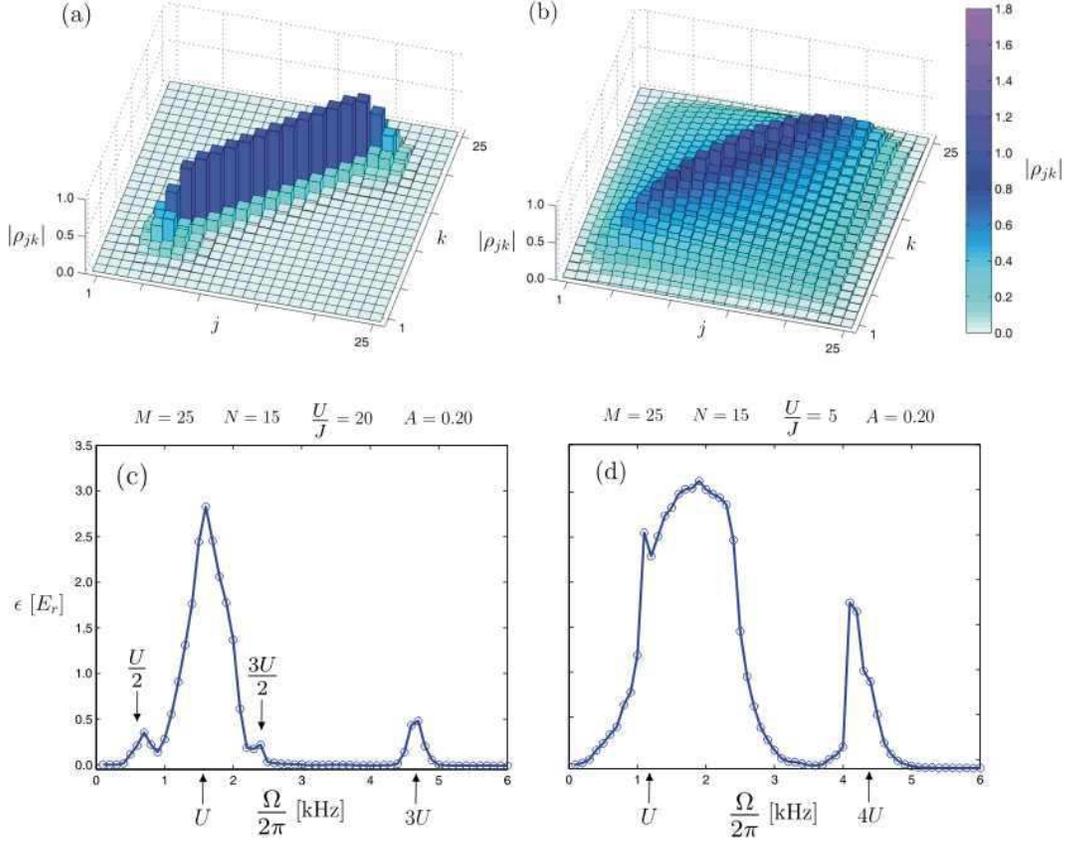}
\caption{The one-particle density matrix $\rho_{jk}$ of the
groundstate for a trapped system with $\omega_T/2\pi=70$ Hz for
(a) the MI regime with $U/J=20$ and (b) the SF regime with
$U/J=5$. The corresponding excitation spectra of the total energy
$\epsilon$ against the modulation frequency $\Omega$ for these
groundstate are shown in (c) and (d)
respectively.}\label{harmonicspec}
\end{center}
\end{figure}

In \fir{harmonicfull} we report the spectra for the system ranging
over SF to MI regimes as before. Again, the essential features of
this plot follow from our earlier discussion. Performing the same
analysis on the evolution of the spectrum demonstrates that
maximum strength of the $3U \rightarrow 4U$ peak has a maximum at
$U/J \sim 8$ as shown in \fir{harmonicanalysis}(a). The trapping
is also seen to introduce peaks and troughs to this curve compared
to the smooth behaviour of the box system. Also, in contrast to
the box both the centre $\Omega_c$ and width $\Omega_w$ of the $U$
resonance show a gradual change from the MI to SF character in
\fir{harmonicanalysis}(b) and (c). This reflects a difference in
the evolution of $U$ resonance, which for the box develops a
visible two-peaked structure in \fir{boxfull}(a), whereas in the
harmonic trap this is washed out into a broader and flatter
response in \fir{harmonicfull}(a).

\begin{figure}[h]
\begin{center}
\includegraphics[width=14cm]{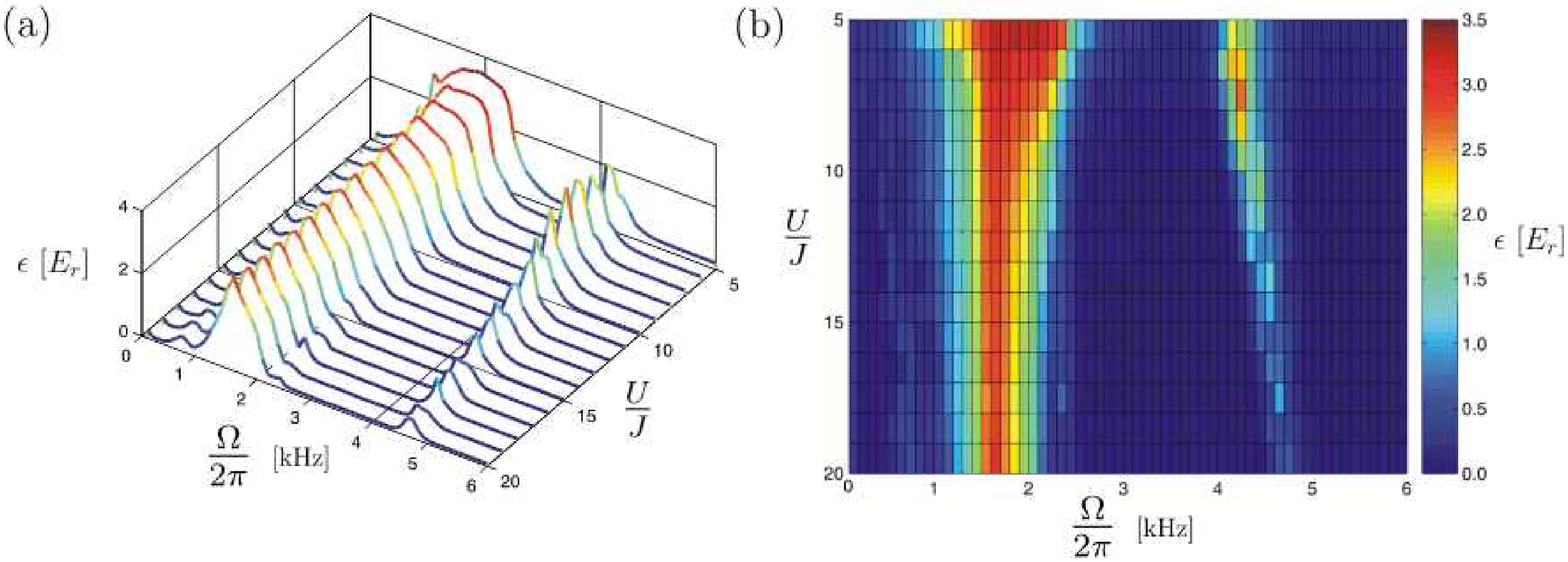}
\caption{(a) The total energy $\epsilon$ absorbed by a $M=25$ and
$N=15$ harmonically trapped system over a sequence of depths,
ranging from the MI to SF regimes, as a function of $\Omega$ for
strong modulations. (b) A colour-map of plot
(a).}\label{harmonicfull}
\end{center}
\end{figure}

\begin{figure}[h]
\begin{center}
\includegraphics[width=14cm]{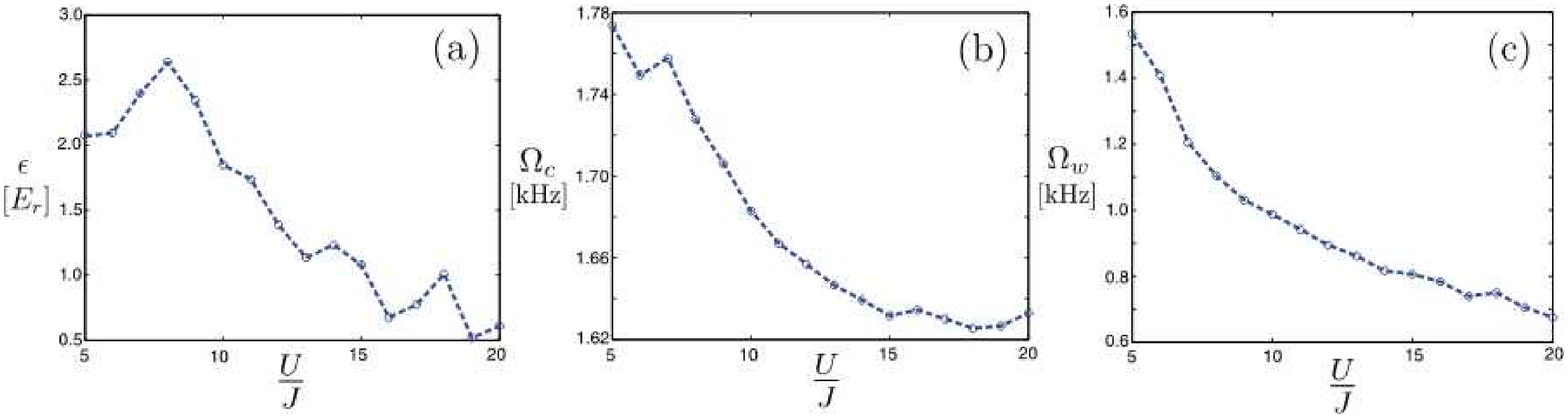}
\caption{(a) The maximum energy $\epsilon$ absorbed for the $3U
\rightarrow 4U$ peak with the lattice depth $U/J$. (b) The centre
position $\Omega_c$, and (c) width $\Omega_w$  of the $U$
resonance with the lattice depth $U/J$. The dotted lines in all
cases is drawn to guide the eye.}\label{harmonicanalysis}
\end{center}
\end{figure}

We have seen that the introduction of harmonic trapping for the
case considered above has not lead to any new features in the
spectrum beyond a small amount of shifting and broadening. One
such feature which might be expected is the emergence of a
resonance at $2U$. However, it is known that the relevant
excitation processes for a $2U$ resonance require either
excitations to already be present in the system due to finite
temperature \cite{Reischl}, or significant inhomogeneity in the
density caused by a trap \cite{Kollath06}. Since the system above
is at $T=0$ and has virtually no incommensurability in the MI,
beyond the small SF lobes at the edges, the lack of a $2U$
resonance is consistent. In the experiment both finite-temperature
and inhomogeneity are expected to make important contributions to
the spectrum. Presently there are still open questions as to which
is the most dominant and what interplay there might be between
these effects. Future studies with the TEBD algorithm, exploiting
recent developments which permit the simulation of master equation
evolution of 1D finite-temperature and dissipative systems
\cite{Zwolak}, could allow both these effects to be treated on an
equal footing.

To demonstrate the contribution of inhomogeneity we have performed
additional calculations again using $M=25$ but with a larger
occupancy of $N=30$ and larger trapping frequency
$\omega_T/2\pi=100$ Hz. In \fir{bigharmonicspec}(a) the
one-particle density matrix of the MI groundstate at $U/J=20$ is
shown and displays the coexistence of significant MI and SF
regions characteristic of trapped systems. Despite using a larger
lattice site dimension $d_s=6$ for this setup we found that
truncation in on-site occupancy limited the smallest $U/J$ from
which we could reliably compute the spectrum to $U/J=9$. The
one-particle density matrix for the $U/J=9$ groundstate is shown
in \fir{bigharmonicspec}(b) and appears to be dominated by a
central SF region. The excitation spectra for these two
groundstates are shown in \fir{bigharmonicspec}(c) and (d)
respectively. For the MI in \fir{bigharmonicspec}(c) a pronounced
$2U$ resonance is seen in addition to the peaks seen in previous
MI spectra. The appearance of a $2U$ peak in the $T=0$ case
examined here arises predominantly from particles in the unit
filled MI shell hopping into the doubly occupied region at the
centre \cite{Kollath06,Stoferle} which is evident in the
one-particle density matrix shown in \fir{bigharmonicspec}(a). For
the shallower depth $U/J=9$ the spectrum in
\fir{bigharmonicspec}(d) exhibits the initial merging of the $U$
and $4U$ peaks with this $2U$ peak. In \fir{bigharmonicfull} the
evolution of the spectrum between these limits is shown. The form
of these spectra is very reminiscent of that shown in the
experiment \cite{Stoferle} and it is likely that a more detailed
analysis of this setup to shallower lattice depths would reveal
the broad SF resonance seen experimentally.

\begin{figure}[h]
\begin{center}
\includegraphics[width=14cm]{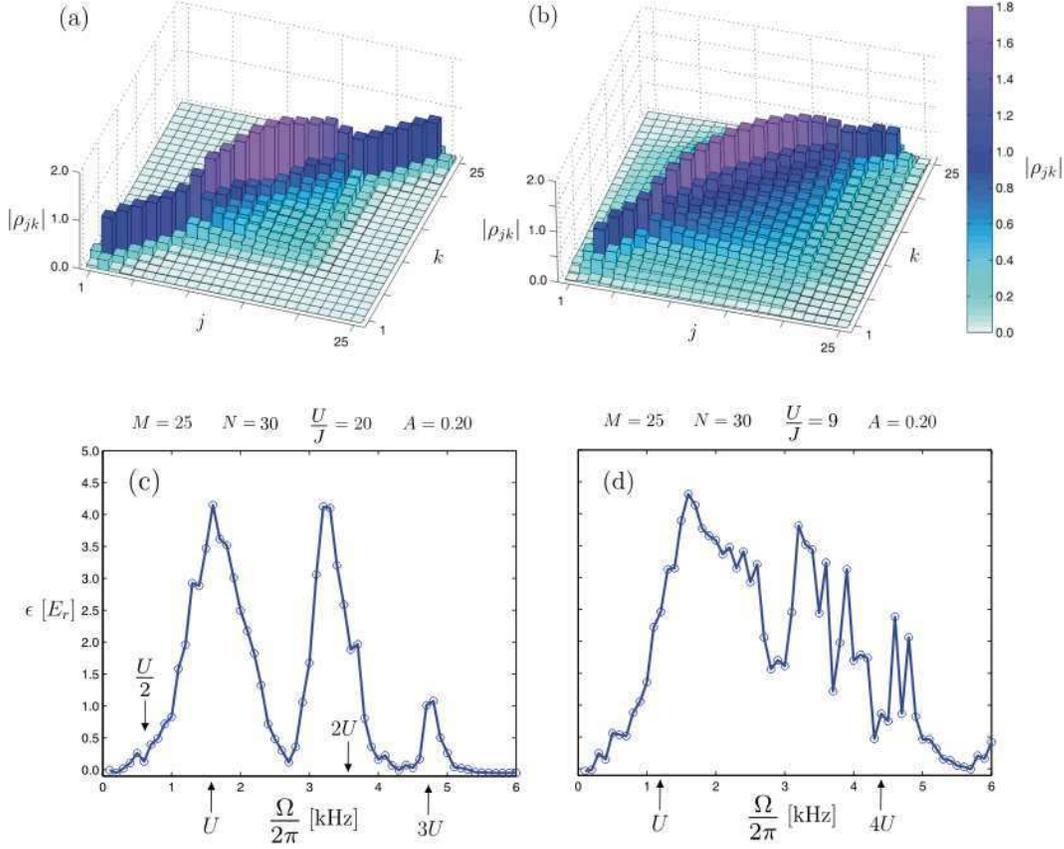}
\caption{The one-particle density matrix $\rho_{jk}$ of the
groundstate for a trapped system with $\omega_T/2\pi=100$ Hz for
(a) the MI regime with $U/J=20$ and (b) an intermediate regime
with $U/J=9$. The corresponding excitation spectra of the total
energy $\epsilon$ against the modulation frequency $\Omega$ for
these groundstate are shown in (c) and (d)
respectively.}\label{bigharmonicspec}
\end{center}
\end{figure}

\begin{figure}[h]
\begin{center}
\includegraphics[width=14cm]{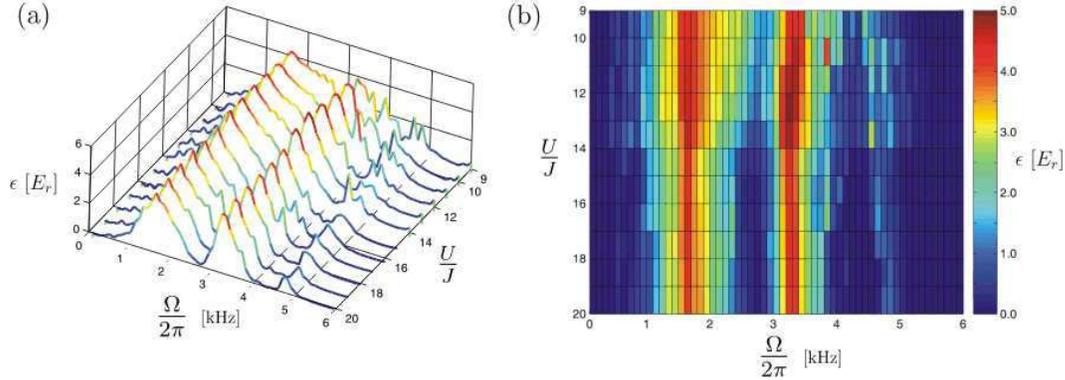}
\caption{(a) The total energy $\epsilon$ absorbed by a $M=25$,
$N=30$ harmonically trapped system over a sequence of depths,
ranging from the MI to SF regimes, as a function of $\Omega$ for
strong modulations. (b) A colour-map of plot
(a).}\label{bigharmonicfull}
\end{center}
\end{figure}

\section{Conclusions}
\label{conc} We have examined in detail the dynamical response at
$T=0$ of the ultracold atoms in an optical lattice subjected to
lattice modulations. We have reported the evolution of the
excitation spectrum with the lattice depth from the SF to MI
regime for small and large box systems, and for large harmonically
trapped systems. For the box system we identified two pronounced
signatures of the transition from the MI to SF regime,
specifically the strength of the $3U \rightarrow 4U$ peak and the
centre of the $U$ resonance. The introduction of trapping for the
case considered here, where the occupancy remains less than or
equal to unity, does not significantly alter the evolution of the
spectrum compared to a box system. While we find that it does wash
out any pronounced changes in the structure of the $U$ resonance
with depth, the strength of the $3U \rightarrow 4U$ peak still
exhibits a signature of increasing SF character.

We also presented calculations showing spectra progressing from
the MI regime for a harmonically trapped system which has a
central region with greater than unit filling. We found that this
changes the structure of the spectrum by introducing a $2U$
resonance and brings our results closer with the experiment.
Direct comparison to the experiment \cite{Stoferle}, however, is
difficult due to the 3D rethermalization performed prior to
measurement, which cannot be simulated with TEBD, and the fact
that the measurement itself was an averaged result over many 1D
systems in parallel with differing total particle number. Despite
this our results point to that fact that the BHM is sufficient to
explain all the features discovered in the experiment and that the
experiment was a clean realization of this model as expected.

\ack S.R.C. and D.J. thank Keith Burnett and Alexander Klein for
stimulating discussions. S.R.C. also thanks Adrian Katian and
Andrew Daley for helpful discussions regarding quantum number
conservation for the TEBD algorithm. This work is supported by the
EPSRC (UK) and the project EP/C51933/1.


\begin{thebibliography}{99}
\bibitem{Shallow}
For the exact system we consider shallow depths which are strictly
beyond the validity of the tight-binding lowest-band BHM. However,
these still provide important information regarding the evolution
of the excitation spectrum at shallow depths.
\end{thebibliography}
\end{document}